# Extremely low linear polarization of comet C/2018 V1 (Machholz–Fujikawa–Iwamoto)

Evgenij Zubko [a,*], Ekaterina Chornaya [b,c], Maxim Zheltobryukhov [c], Alexey Matkin [c], Oleksandra V. Ivanova [d,e,f], Dennis Bodewits [g], Anton Kochergin [b,c], Gennady Kornienko [c], Igor Luk'yanyk [d], Dean C. Hines [h], Gorden Videen [i,j]

[a] *Humanitas College, Kyung Hee University, 1732, Deogyeong-daero, Giheung-gu, Yongin-si, Gyeonggi-do 17104, South Korea*
[b] *Far Eastern Federal University, 8 Sukhanova St., Vladivostok 690950, Russia*
[c] *Institute of Applied Astronomy of RAS, 10 Kutuzova Emb., Saint-Petersburg 191187, Russia*
[d] *Astronomical Observatory, Taras Shevchenko National University of Kyiv, 3 Observatorna St., Kyiv, 04053, Ukraine*
[e] *Astronomical Institute of the Slovak Academy of Sciences, SK-05960 Tatranská Lomnica, Slovak Republic*
[f] *Main Astronomical Observatory of National Academy of Sciences, 27 Akademika Zabolotnoho St., Kyiv, 03143, Ukraine*
[g] *Auburn University, Physics Department, Auburn, AL 36849-5319, USA*
[h] *Space Telescope Science Institute, 3700 San Martin Drive, Baltimore, MD 21218, USA*
[i] *Space Science Institute, 4750 Walnut Street, Boulder Suite 205, CO 80301, USA*
[j] *Department of Astronomy and Space Science, Kyung Hee University, 1732, Deogyeong-daero, Giheung-gu, Yongin-si, Gyeonggi-do 17104, South Korea*

We measured the degree of linear polarization $P$ of comet C/2018 V1 (Machholz-Fujikawa-Iwamoto) with the broadband Johnson $V$ filter in mid-November of 2018. Within a radius of $\rho \approx 17{,}000$ km of the inner coma, we detected an extremely low linear polarization at phase angles $\alpha \approx 83°$–$91.2°$ and constrained the polarization maximum to $P_{max} \approx (6.8 \pm 1.8)\%$. This is the lowest $P_{max}$ ever measured in a comet. Using model agglomerated debris particles, we reproduced the polarimetric response of comet C/2018 V1. Four retrieved refractive indices closely match what was experimentally found in Mg-rich silicates with little or no iron content. Moreover, the size distribution of the agglomerated debris particles appears in good quantitative agreement with the *in situ* findings of comet 1P/Halley. The dust model of polarization of comet C/2018 V1 suggests a strongly negative polarization with amplitude $|P_{min}| \approx 5\%$–$7\%$; whereas, an interpretation based on gaseous emission requires no negative polarization at small phase angles. This dramatic difference could be used to discriminate gaseous- emission and dust explanations in low-$P_{max}$ comets in future.

## Introduction

When initially unpolarized solar radiation is scattered from the gas and dust surrounding a comet, it acquires a partial linear polarization. Its quantity is described in terms of the degree of linear polarization $P$:

$$P = -\frac{Q}{I} = \frac{I_\perp - I_\|}{I_\perp + I_\|} \qquad (1)$$

Here, $Q$ and $I$ denote the Stokes parameters (Bohren and Huffman, 1983); $I_\perp$ and $I_\|$ are the intensity of components of the scattered light that are polarized perpendicularly to the scattering plane and within the scattering plane, respectively. The polarization $P$ is often expressed in percent. Clearly, $P$ in Eq. (1) is a sign-dependent characteristic whose negative value corresponds to the case of $I_\perp < I_\|$ and positive value to $I_\perp > I_\|$.

The degree of linear polarization $P$ varies significantly with the observation/illumination geometry of the comet. For instance, at small phase angles ($\alpha < 25°$), the polarization is predominantly negative. At larger phase angles ($\alpha > 25°$), $P$ takes on positive values (e.g., Chernova et al., 1993; Levasseur-Regourd et al., 1996). In both domains of $\alpha$, the angular profile of polarization tends to appear in a form of a simple, bell-shape branch. While the branch of negative polarization is characterized by a minimum polarization $P_{min}$ at phase angle $\alpha_{min}$, the positive polarization branch reaches its maximum $P_{max}$ at a phase angle $\alpha_{max}$. It is interesting to note that laboratory measurements of light scattering by single cometary-analog particles and meteorite particles demonstrate qualitatively similar behavior (e.g., Muñoz et al., 2000; Frattin et al., 2019), although specific values of $P_{min}$, $α_{min}$, $P_{max}$, and $α_{max}$ may significantly differ from one target to another.

Comets demonstrate a significant dispersion of their $P_{max}$ (Chernova et al., 1993) and were initially classified as high-$P_{max}$ or low-$P_{max}$ (Levasseur-Regourd et al., 1996). A third class was established based on the very high $P_{max}$, observed in comet C/1995 O1 (Hale-Bopp) (Hadamcik and Levasseur-Regourd, 2003a). However, these segregated classes seemingly result from limited data, as a recent analysis of the extended dataset in Zubko et al. (2016) suggests a continuous transition from the low-$P_{max}$ comets to the high-$P_{max}$ comets, and further to C/1995 O1 (Hale-Bopp). This can be seen in Fig. 1 where we reproduce a compilation of polarimetric measurements of 23 different comets considered by Zubko et al. (2016).

The dispersion was initially interpreted in terms of the different relative contributions of the gaseous emission into light scattering from cometary coma (Chernova et al., 1993) because such emission is long known to be highly depolarized. For instance, complex gaseous molecules have $P_{max}$ = 7.7% (e.g., Le Borgne et al., 1987 and therein). A few years later, Levasseur-Regourd et al. (1996) attributed the dispersion of $P_{max}$ in comets to different microphysical properties of their dust. Until recently, both these interpretations were considered to be possible (e.g., Kiselev et al., 2015). Zubko et al. (2016) demonstrated that the gaseous emission model is inconsistent with several existing observations of comets, and they developed a simple model that can satisfactorily reproduce the dispersion of $P_{max}$ in comets using rigorous simulations of light scattering by particles with highly irregular and fluffy morphology, and that the different polarizations are a result of having different material compositions. The modeling was based on an assumption that the light-scattering signal was dominated by two different types of particles, one with low (Im($m$) ≤ 0.01) and another with high (Im($m$) > 0.4) imaginary part of complex refractive index. While the former component is consistent with Mg-rich silicates (Dorschner et al., 1995), the latter one appears consistent with some organics (Jenniskens, 1993) and amorphous carbon (Duley, 1984). The presence of these species in comets is well established (e.g., Fomenkova et al., 1992; Cochran et al., 2015). The dispersion of polarization in comets can be explained then by variation of a sole parameter, the volume ratio of the weakly absorbing particles and highly absorbing particles. Within this two-component framework, the coma in comets with low $P_{max}$ predominantly consists of Mg-rich silicates; whereas, in the high-$P_{max}$ comets, it is abundant with carbonaceous materials.

What emerges from Fig. 1 is a disparity between the number of high-$P_{max}$ comets and low-$P_{max}$ comets. There are well over a dozen representatives of $P_{max}$ > 20%, and only a few comets with $P_{max}$ ≤ 10%. There are three comets, 23P/Brorsen–Metcalf, 27P/Crommelin, and C/1975 N1 (Kobayashi–Berger–Milon) whose polarimetric response appears with confidence at the bottom limit in Fig. 1; all these observations were reported in Chernova et al. (1993). One more comet, C/1996 Q1 (Tabur), had $P ≈ 13–15\%$ at $α ≈ 85.5°$ during a period of normal activity. However, for no obvious reason, this comet suddenly lost its activity and coma. This process was accompanied with a decrease of polarization to $P ≈ 10\%$ at $α ≈ 85°$ (Kikuchi, 2006). Thus, strictly speaking, only three comets revealed a persistent maximum of polarization $P_{max}$ ≤ 10%. It is worth noting that the last of them was observed about thirty years ago, in 1989. In this manuscript we report our polarimetric observations of a newly discovered comet C/2018 V1 (Machholz–Fujikawa–Iwamoto) at large phase angles, $α ≈ 83°–91.2°$ This is only the fourth example of a comet with such a low maximum of positive polarization (Fig. 1).

**Observations and data processing**

We conducted polarimetric observations of comet C/2018 V1 shortly after its discovery on November 7, 2018. The comet passed perihelion at 0.387 au on December 3, 2018. Weather conditions were favorable on three dates, November 16, 17, and 18. However, on the first date, the observations were interrupted by clouds when only two series of images were taken. Nevertheless, they appear in very good accordance with results obtained on the other two nights and, therefore, we include them into the analysis. Observations were made with the 22-cm telescope located at the Ussuriysk Astrophysical Observatory (code C15), which operates within the International Scientific Optical Network (ISON). We used CCD detector FLI ProLine PL4301E that has a resolution of 2084 × 2084 pixels and pixel size of 24 μm. The field of view of the CCD detector is 326 × 326 arcmin with angular resolution of 9.39 × 9.39 arcsec per pixel. The Johnson $V$ filter was used for polarimetric observation of the comet. We use a dichroic polarization filter (analyzer), which rotated sequentially through three fixed position angles 0°, +60°, and +120° The details of observations are summarized in Table 1. Morphology of the C/2018 V1 coma can be seen in Fig. 2.

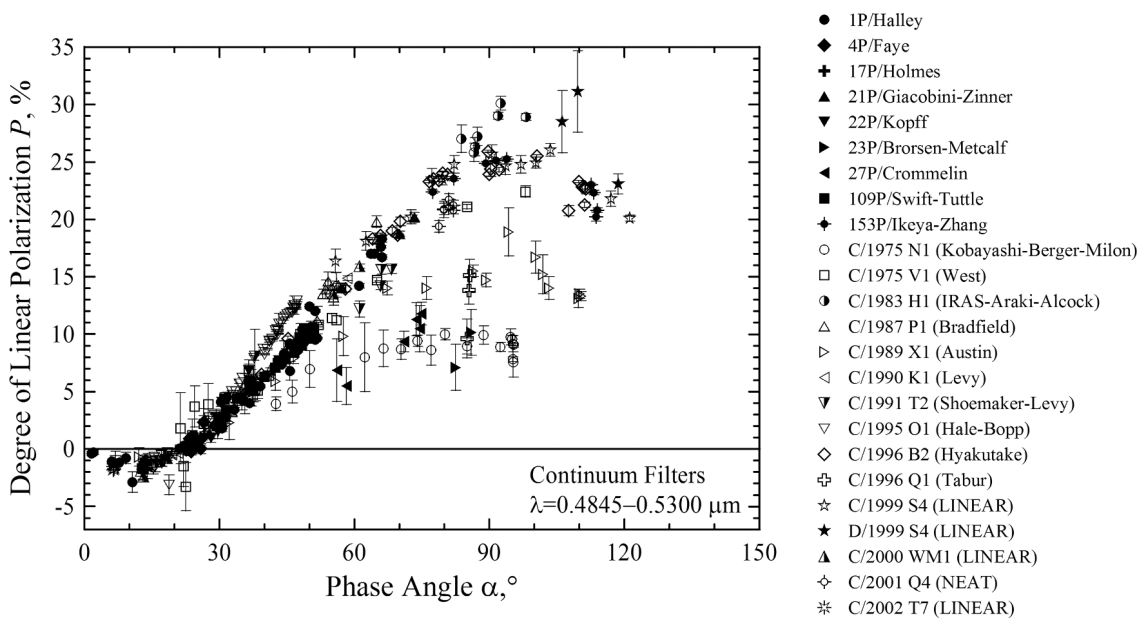

**Fig. 1.** Degree of linear polarization $P$ measured with continuum filters in the blue-green part of the spectrum in 23 different comets as a function of phase angle α. Compilation is adapted from Zubko et al. (2016), where references and some details of the observation of each specific comet are given.

The obtained data have been processed using the Image Reduction and Analysis Facility (IRAF) software system for reduction and analysis of astronomical data. Our processing includes bias subtraction, removal of cosmic-ray events and flat-field correction. Flat-field correction was constructed based on images of the twilight sky. We used a procedure that retrieves a histogram of counts of the sky background. This allows us to determine accurately its maximum level and then subtract it from every image. For calibration purposes, we observed polarized and non-polarized standard stars from the lists of Hsu and Breger (1982), Schmidt et al. (1992), and Heiles (2000). Our instrumental polarization is about 0.3 percent under good atmospheric conditions.

Using the images obtained at three evenly distributed orientations of the analyzer, 0°, 60°, and 120° measured from the scattering plane, we compute the corresponding fluxes from the inner coma with $\rho \approx 17{,}000$ km: $F_0$, $F_{60}$, and $F_{120}$. These values can be translated into the $I$, $Q$, and $U$ Stokes parameters using the Fesenkov formulae (e.g., Fessenkoff, 1935; Hines et al., 2014) as follows:

$$I = \frac{2}{3}(F_0 + F_{60} + F_{120}),$$
$$Q = \frac{2}{3}(2F_0 - F_{60} - F_{120}), \quad (2)$$
$$U = \frac{2}{\sqrt{3}}(F_{60} - F_{120}).$$

In what follows, the Stokes parameters are utilized to infer the total degree of linear polarization $P_{\text{total}}$ and the angle $\theta$ between the semi-major axis of the vibration ellipse and the scattering plane (Bohren and Huffman, 1983):

$$P_{\text{total}} = \frac{\sqrt{Q^2 + U^2}}{I} \text{ and } \tan(2\theta) = \frac{U}{Q} \quad (3)$$

Table 2 presents $P_{\text{total}}$ and $\theta$ with their error bars as measured in comet C/2018 V1. As one can see from values of $\theta$, the plane of linear polarization appears to be almost perpendicular to the scattering plane. As was noticed in Section 1, this is a common feature of comets observed at large phase angles. However, polarimetric studies of comets refer typically to the sign-dependent definition degree of linear polarization with Eq. (1). Transformation of $P_{\text{total}}$ and $\theta$ in Eq. (3) to $P$ in Eq. (1) takes on a simple form (Zubko and Chornaya, 2019):

$$P = -P_{\text{total}}\cos(2\theta). \quad (4)$$

The results of this transformation are shown in the last column in Table 2. It is worth noting that the error bars in $P$ are twice larger than in $P_{\text{total}}$ that results from error bars accompanying measurements of the angle $\theta$. It is finally worth noting that the uncertainty in our measurements of $P$ in comet C/2018 V1 appear consistent with what emerged from studies of other comets (see, Fig. 3).

## 3. Results and discussion

In Fig. 3 we plot the degree of linear polarization $P$ as a function of phase angle α in comet C/2018 V1 compared with three other comets having comparably low positive polarization at side scattering (adapted from Chernova et al., 1993). As one can see, the polarization of C/2018 V1 appears at the bottom limit even in this featured group. Interestingly, the polarimetric response in C/2018 V1 appears accidently consistent with the angular profile of polarization of fluorescence from complex molecules and ions (shown with the dashed line), like $C_2$ molecules (Le Borgne et al., 1987). Such resemblance of the low-$P_{\max}$ comets was attributed by Chernova et al. (1993) to gaseous emission. We refer to Zubko et al. (2016) for a detailed discussion of the inconsistencies that can result from this assumption. Here we only note the absence of the negative polarization branch that occurs near backscattering angles for the gaseous emission. To our knowledge this was never found in the coma-integrated polarimetric measurements of comets with the narrowband continuum filters, like those used in Chernova et al. (1993), or broadband filter, the one that we use in this work. Some features of the cometary coma, such as jets or arcs, indeed reveal essentially positive polarization at small phase angles. It is significant, however, that their relative contribution to the aperture-integrated response does not eliminate the negative polarization branch (e.g., Hadamcik and Levasseur-Regourd, 2003b; Hines et al., 2014).

In the broadband Johnson $V$ filter that we used in our observations, the strongest gaseous contamination might arise from the $C_2$ gaseous emission near $\lambda = 0.514$ μm. The narrowband continuum filters in the blue-green part of spectrum that are used in Chernova et al. (1993) and in many other polarimetric observations of comets are much better isolated from the $C_2$ emission (e.g., Farnham et al., 2000). Owing to their different spatial distribution of gas and dust, gas contributes little to the light scattering near the nucleus (Jewitt et al., 1982; Picazzio et al., 2019). As a consequence, the difference in polarization simultaneously measured in a comet with the broadband $V$ filter and with a narrowband continuum filter whose bandpass is embraced by the $V$ filter hardly exceeds the corresponding error bars (Zubko et al., 2014; Ivanova et al., 2017).

The aperture size used to image the coma strongly influences the relative contribution of the $C_2$ emission. The dust in the coma typically peaks on the optocenter with a $1/r$ distribution, while the gas distribution, in particular that of a fragment species that is formed in the coma, has a much flatter distribution (cf. Combi and Fink, 1997). The $C_2$ molecule does not directly sublimate from the nucleus but is likely a dissociation product of multiple larger molecular and/or cometary CHON grains (A'Hearn et al., 1995; Combi and Fink, 1997; Feldman et al., 2004; McKay et al., 2014). The number of $C_2$ molecules in a coma increases as the number of their parent molecules decreases with $N = N_0 \exp(-d/R)$, where $N_0$ is the initial number of parent molecules, $R$ is their (empirical) Haser scale length, and $d$ is the distance to the nucleus. Evidently, $N = N_0$ at $d \to 0$ and $N = 0$ at $d \to \infty$. We note that a multistep Haser model can also take the destruction of the fragment species into account (c.f. Festou, 1981), but given that this mostly occurs outside of our field of view, we do not address this phenomenon here.

On November 16–18 of 2018, comet C/2018 V1 was at a heliocentric distance $r_h \approx 0.55$–0.60 au In similar circumstances, at $r_h \approx 0.7$ au, Fink et al. (1991) measured profiles of $C_2$ based on observations of comet Halley and determined empirical Haser scale lengths of $\sim 10^5$ km. These values depend on the heliocentric distance, solar activity cycle, and the source of the $C_2$ molecule, but are likely of similar order for C/2018 V1 at 0.55–0.6 au Scaling by $r_h^2$, we find a Haser scale length $R$ of $6 \times 10^4$ km. That means that only 25% of the $C_2$ progenitors dissociate within an aperture of $\rho = 1.7 \times 10^4$ km:

$$N/N_0 = \exp(-\rho/R) = \exp(-1.7 \times 10^4 / 6 \times 10^4) \sim 75\%. \quad (5)$$

**Table 1**
Log of observations of comet C/2018 V1 in November of 2018.

| UT date, 2018–11 | $r_h$, au | Δ, au | α, ° | ρ, km | ρ, arcsec | Airmass | Frames | Exposure, sec |
| --- | --- | --- | --- | --- | --- | --- | --- | --- |
| 16.829 | 0.587 | 0.870 | 83.0 | 17,800 | 18.8 | 4.630 | 2 × 3 | 30 |
| 17.840 | 0.569 | 0.839 | 87.0 | 17,100 | 19.7 | 5.002 | 10 × 3 | 30 |
| 18.833 | 0.551 | 0.809 | 91.2 | 16,500 | 20.7 | 5.528 | 11 × 3 | 30 |

ρ – the aperture radius.

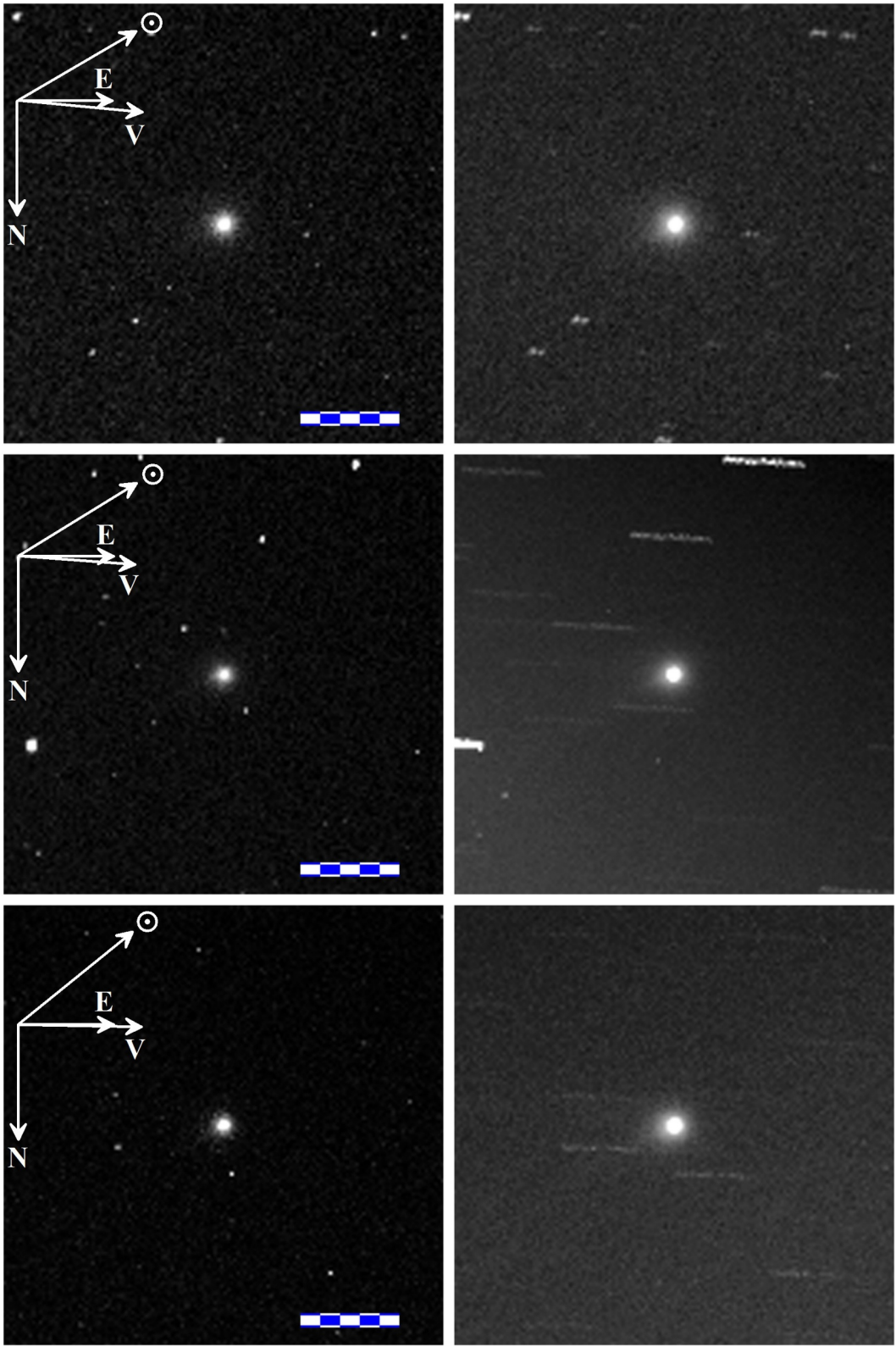

**Fig. 2.** Images of coma of C/2018 V1 obtained on November 16, 17, and 18 of 2018 (from top to bottom). Examples of single images are shown on left; whereas, a stack of all images obtained on the same night—on right. Single scale mark corresponds to diameter of the circular aperture used in integration of the polarimetric response.

**Table 2**
Polarization of comet C/2018 V1 in November of 2018.

| UT date, 2018–11 | α, ° | $P_{total}$,% | θ, ° | P,% |
|---|---|---|---|---|
| 16.829 | 83.0 | 7.03 ± 1.06 | 101.9 ± 7.2 | 6.34 ± 1.65 |
| 17.840 | 87.0 | 7.90 ± 1.10 | 106.7 ± 6.5 | 6.56 ± 1.88 |
| 18.833 | 91.2 | 8.15 ± 1.09 | 106.7 ± 5.7 | 6.80 ± 1.78 |

$C_2$ may be produced by the dissociation of one or more parent molecules, or by the decay of CHON particles. As will be shown below, the coma of C/2018 V1 is highly depleted of highly absorbing dust that is indicative of the presence of CHON particles.

As mentioned in Section 1, the dispersion of $P_{max}$ in comets can be explained in terms of different light-scattering properties of their dust. In particular, Zubko et al. (2016) developed a model of light scattering by dusty coma that can satisfactorily reproduce multi-wavelength polarimetric observations of numerous comets. It is based on the simulation of the irregularly shaped cometary dust with model *agglomerated debris particles*. These particles have a highly disordered, fluffy morphology with a packing density of their constituent material being about 0.236 with regard to a circumscribing sphere. Six examples of agglomerated debris particles are shown in the top panels of Fig. 4 with three micron-sized dust particles (bottom) that were sampled by *Rosetta* in the vicinity of the 67P/Churyumov–Gerasimenko nucleus and studied with the *Micro-Imaging Dust Analysis System* (MIDAS). These images are adapted from Bentley et al. (2016). As one can see in Fig. 4, the shapes of agglomerated debris particles resemble those of cometary dust in comet 67P/Churyumov–Gerasimenko.

Unfortunately, the other *Rosetta* instruments were insensitive to submicron- and micron-sized particles (e.g., Güttler et al., 2019).

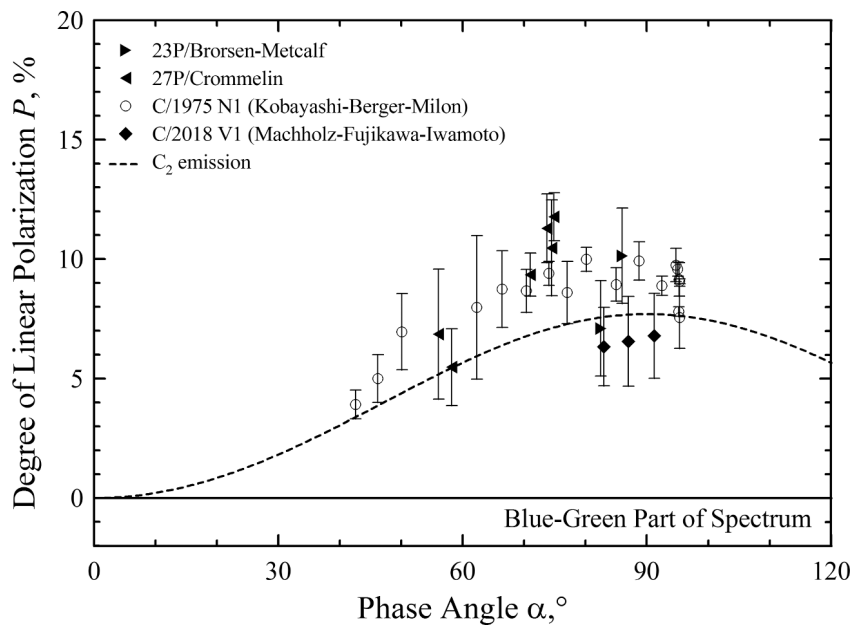

**Fig. 3.** Degree of linear polarization P in comet C/2018 V1 as a function of phase angle α versus three other comets with low $P_{max}$ reported in Chernova et al. (1993). The dashed line shows the angular profile of polarization of the $C_2$ emission (e.g., Le Borgne et al., 1987).

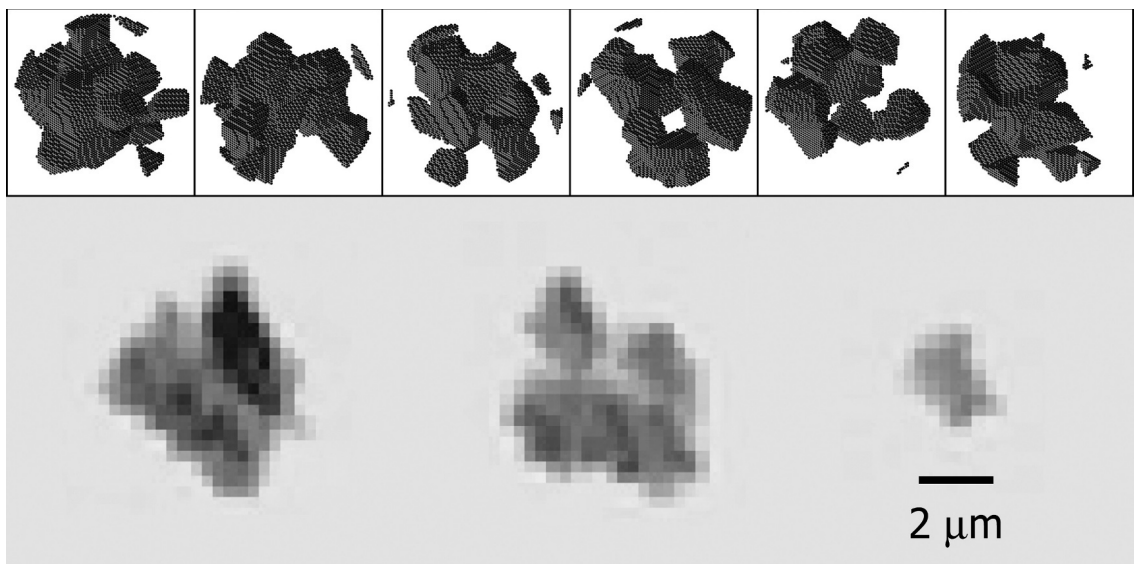

**Fig. 4.** Top: Images of six samples of irregularly shaped *agglomerated debris particles* that are used in modeling of polarization of C/2018 V1. Bottom: three micron-sized dust particles sampled by *Rosetta* in the innermost coma of 67P/Churyumov–Gerasimenko (adapted from Bentley et al., 2016).

Therefore, we constrain the physical and chemical properties of such small cometary dust particles using results obtained *in situ* in other comets. For instance, the shape of the agglomerated debris particles also agrees with what was found in dust particles of comet 26P/Grigg–Skjellerup captured in the stratosphere (Busemann et al., 2009) as well as with the shape inferred from micron-sized craters produced by dust from comet 81P/Wild 2 in Al foil of the *Stardust* space probe (Hörz et al., 2006).

Under the assumption of ROCK (silicate) and/or CHON (carbonaceous) composition, similar to dust in comet 1P/Halley (e.g., Fomenkova et al., 1992), the bulk density of the agglomerated debris particles spans the range from $0.35\,\mathrm{g/cm^3}$ to $0.83\,\mathrm{g/cm^3}$ that appears in good accordance with the *Stardust* findings of comet 81P/Wild 2 (Hörz et al., 2006) as well as interplanetary dust particles (IDPs) (Flynn and Sutton, 1991).

Light scattering by agglomerated debris particles is computed with the *discrete dipole approximation* (DDA), a flexible numerical technique that allows rigorous consideration of the interaction of electromagnetic waves with an arbitrary-shape target whose size is comparable to the wavelength of the incident light $\lambda$ (e.g., review by Yurkin and Hoekstra, 2007). Using our own implementation of the DDA (e.g., Zubko et al., 2016), we compute the light scattering by agglomerated debris particles at various values of complex refractive index $m$ and size parameter $x = 2\pi r/\lambda$, where $r$ is the radius of the circumscribing sphere of the agglomerated debris particles. Almost 50 different values of complex refractive index $m$ have been investigated to date. Each refractive index $m$ was studied over wide range of $x = 1$–32 (50 in the case of optically soft materials with $\mathrm{Re}(m) < 1.4$). At each pair of $m$ and $x$, the light-scattering properties are averaged over randomly generated shapes and orientations of the agglomerated debris particles. A minimum of 500 samples are typically taken into account to provide a statistically reliable result. We refer to Zubko et al. (2016) for more technical details on the generation of the agglomerated debris particles and DDA computations of their light-scattering response.

We consider agglomerated debris particles obeying a power-law size distribution $r^{-n}$, where the power index ranges from $n = 1.5$ to 3. This range of $n$ was detected *in situ* in submicron- and micron-sized dust particles of comet 1P/Halley (Mazets et al., 1986). However, it also embraces value $n \approx 2.89$ that was inferred from size distribution of micrometer craters in the *Stardust* sampling module after its close encounter with comet 81P/Wild 2 (Price et al., 2010). One needs to note that every agglomerated debris particle consists of a bunch of chunks, which are polydisperse in size. Moreover, each chunk of material is made of a number of closely packed small cells. Such discrete morphology appears in good accordance with what was found in the 67P dust particles (see Fig. 4 and Bentley et al., 2016). However, the power-law size distribution refers to the population of the agglomerated debris particles, but not their fragments.

The dispersion of positive polarization in comets was reproduced by Zubko et al. (2016) with a mixture of two types of agglomerated debris particles. One of them includes, for instance, Mg-rich silicates ($\mathrm{Im}(m) \leq 0.01$; see, e.g., Dorschner et al., 1995), and the other consists of organics and/or amorphous carbon ($\mathrm{Im}(m) > 0.4$; see, e.g., Jenniskens, 1993; Duley, 1984). The phase dependence of the degree of linear polarization in a given comet can be fitted by varying a single free parameter: the volume ratio of the two different species, which could correspond, for instance, to the ratio of Mg-rich silicate particles to carbonaceous particles. We note that the photometric and polarimetric observations of comets often reveal at least two types of dust in their coma (e.g., Zubko et al., 2012; 2013; 2014; 2015; Ivanova et al., 2017; Markkanen et al., 2018; Luk'yanyk et al., 2019). Only in rare instances, can observations be quantitatively reproduced under the assumption of a single component of their dust. (Picazzio et al., 2019).

Within the two-component framework, low $P_{\max}$ in comets C/1975 N1 (Kobayashi–Berger–Milon) and 23P/Brorsen–Metcalf were interpreted to have a high relative abundance of Mg-rich silicate particles, 82%–95% by volume (Zubko et al., 2016). However, as follows from Fig. 3, the polarization in C/2018 V1 tends to take even lower values compared to these two comets, suggesting an even greater relative abundance of the non-absorbing component. Therefore, we search for the best fit to the degree of linear polarization in comet C/2018 V1 without absorbing particles. We consider a few plausible assumptions on their real part of refractive indices $\mathrm{Re}(m)$. Fig. 5 shows results of modeling obtained at $\mathrm{Re}(m) = 1.5$ (top), 1.6 (middle), and 1.7 (bottom). In each case, we include different values of the imaginary part of refractive index $\mathrm{Im}(m)$ that is representative of small iron content in the Mg-rich silicates (Dorschner et al., 1995).

In Fig. 5 we plot only results obtained when the power index is constrained to the range $1.5 \leq n \leq 3$. This supplementary condition yields only one curve that corresponds to zero material absorption (i.e., $\mathrm{Im}(m) = 0$ and a power index $n = 1.7$), shown on the top panel (Re $(m) = 1.5$). An increase of $\mathrm{Re}(m)$ makes it possible to fit the data if we include a small amount of absorption. For instance, at $\mathrm{Re}(m) = 1.6$ (middle panel), observations are equally well reproduced with $\mathrm{Im}(m) = 0.0005$ and $n = 2.2$ and with $\mathrm{Im}(m) = 0.01$ and $n = 1.7$. A further increase of $\mathrm{Re}(m)$ to 1.7 can be used to fit the polarization in C/2018 V1 when $\mathrm{Im}(m) \leq 0.02$ (bottom panel). Clearly, the growth of $\mathrm{Re}(m)$ extends the range of $\mathrm{Im}(m)$ and increases $n$. It is worth noting that the fits obtained at the same $\mathrm{Re}(m)$ tend to converge.

At small phase angles, all the curves shown in Fig. 5 reveal a deep branch of negative polarization, whose amplitude exceeds 5%. This branch is a few times deeper than in comets with high $P_{\max}$ (e.g., Chernova et al., 1993; Levasseur-Regourd et al., 1996). Similar enhanced negative polarization was previously found in the *circumnucleus halo* in some comets (Hadamcik and Levasseur-Regourd, 2003b). It was shown that this phenomenon arises from a local domination of Mg-rich silicate particles in the proximity of the nucleus and, moreover, that the existence of the circumnucleus halo is consistent with the *Stardust* findings in comet 81P/Wild 2 (Zubko et al., 2012). We note also that *in situ* polarimetric measurements of comets 1P/Halley and 26P/Grigg-Skjellerup conducted by the *Giotto* space probe reveal a very low positive polarization at side scattering in the vicinity of their nuclei (Levasseur-Regourd et al., 1999; McBride et al., 1997). This feature correlates with the strong negative polarization in the circumnucleus halo as both phenomena result from cometary dust having weak material absorption (Zubko et al., 2013). It would appear that this also is the case for the chemical composition of dust in comet C/2018 V1. However, while the typical radius of the circumnucleus halo is only 1000–2000 km (Hadamcik and Levasseur-Regourd, 2003b), in this case, it is the entire coma of comet C/2018 V1 that has this strong feature out to a radius of $\sim$17,000 km.

The four fits obtained at $m = 1.6 + 0.0005i$, $1.6 + 0.01i$, $1.7 + 0.01i$, and $1.7 + 0.02i$ suggest particles composed of Mg-rich silicates with little or no iron content (Dorschner et al., 1995). Such domination of silicates in the C/2018 V1 coma could imply the presence of the *10-μm silicate feature*, which manifests itself in a thermal emission spectrum at wavelengths $\lambda \approx 9$–12 μm (e.g., Hanner and Bradley, 2004; Kimura et al., 2009). To our knowledge, no mid-IR observations of comet C/2018 V1 have been reported in the literature yet. They plausibly were not conducted at all in view of the short time period between discovery of this comet and its closest approach with Earth ($\sim$10 days). We stress, however, that mid-IR spectra have been measured in two out of three of the lowest-$P_{\max}$ comets known prior C/2018 V1. Results obtained in those two comets are somewhat intricate. While comet 23P/Brorsen–Metcalf did reveal a strong 10-μm silicate feature that unambiguously suggests the presence of silicates in its coma, comet C/1975 N1 (Kobayashi–Berger–Milon) did not (Gehrz and Ney, 1992). The latter fact does not immediately imply the absence of silicate particles in the coma as the strength of the 10-μm silicate feature is governed not only by the presence of silicate material, but also, by the size distribution of the silicate particles. For instance, if there is a considerable amount of compact silicate particles whose size significantly exceeds 1 μm, the coma may not produce the feature (Hanner and Bradley, 2004). On the other hand, such super-micron silicate particles do not affect the light-scattering response in the visible, as it is largely produced by particles whose size is comparable to the wavelength of the incident light (e.g., A'Hearn et al., 1995).

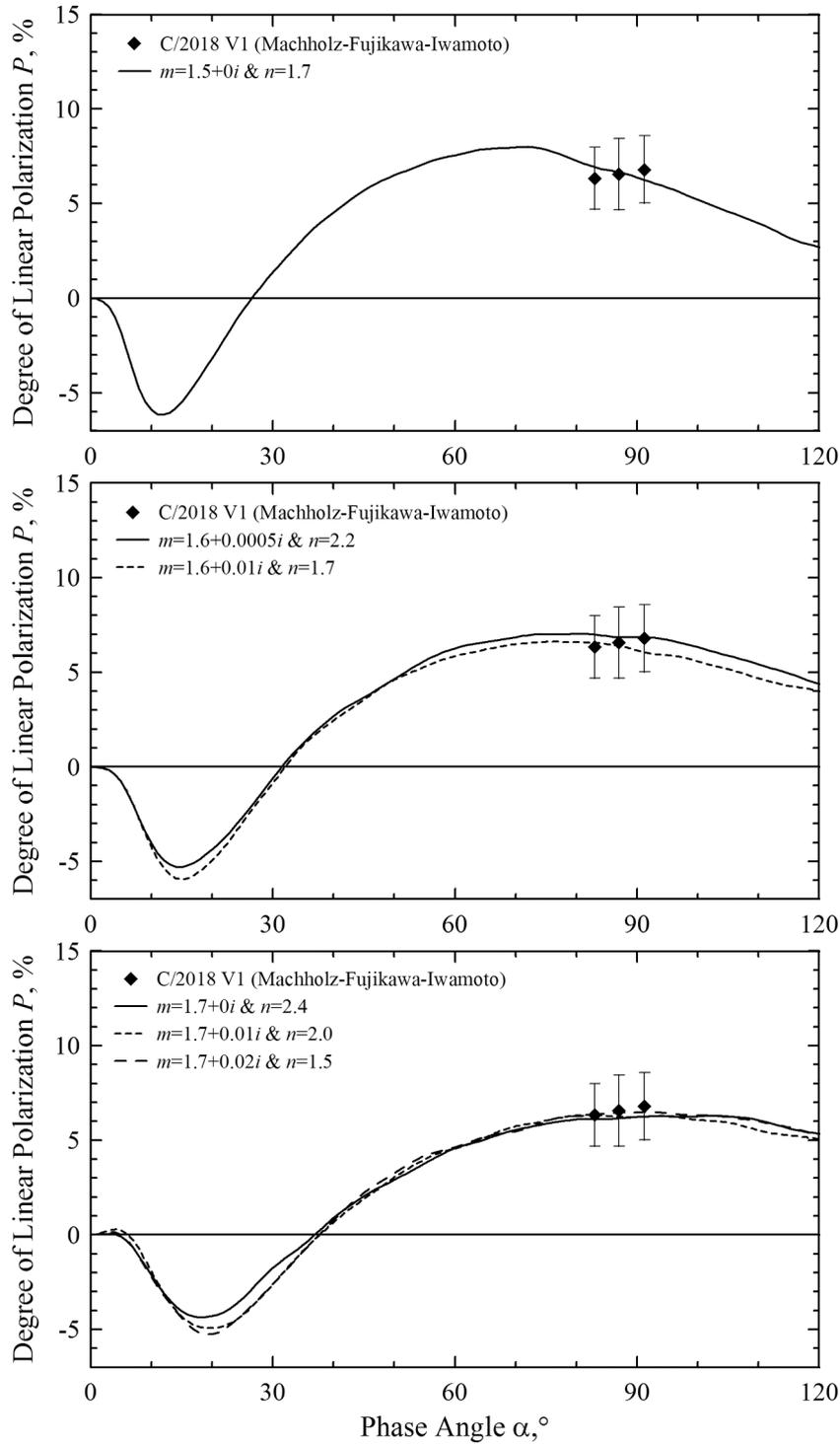

**Fig. 5.** Modeling of polarization in comet C/2018 V1 with the agglomerated debris particles. Panels from top to bottom demonstrate results obtained with the real part of refractive index Re($m$) = 1.5, 1.6, and 1.7, respectively.

### Conclusion remarks

In mid-November of 2018, we measured the degree of linear polarization $P$ in comet C/2018 V1 (Machholz–Fujikawa–Iwamoto) using the broadband Johnson $V$ filter at large phase angles, $\alpha \approx 83°$–$91.2°$, where the maximum value of polarization $P_{max}$ is typically attained in comets. We infer a constraint on $P_{max} \approx (6.8 \pm 1.8)$%, which is the smallest $P_{max}$ ever measured in cometary comae. Comparable $P_{max}$ was previously found only in three other comets, the last of which was some three decades ago. Because of the relatively large error bars in the polarization measurements, low polarization in comet C/2018 V1 could be formally consistent with $C_2$ molecular emission. The total linear polarization $P_{total}$ in C/2018 V1 is lower than the polarization of the $C_2$ gaseous emission at the same phase angles. Moreover, $C_2$ molecules result from dissociation of more complex organic molecules. This process takes a relatively long time during which the parent molecules escape to a significant distance from the nucleus. However, we measure polarization in the coma whose radius is of only ~17,000 km. These facts suggest that the low positive polarization in comet C/2018 V1 is governed by dust in its coma.

We demonstrate that model *agglomerated debris particles* can satisfactorily reproduce the polarimetric response in comet C/2018 V1 using six different particle distributions, all of which have weak material absorption. Four of the refractive indices, $m = 1.6 + 0.0005i$, $1.6 + 0.01i$, $1.7 + 0.01i$, and $1.7 + 0.02i$, are similar to laboratory measurements of refractive index of Mg-rich silicate in the visible (Dorschner et al., 1995). The best fits to polarization in C/2018 V1 were obtained using the power index $n = 1.5 – 2.2$, that also are in good quantitative agreement with *in situ* measurements of the size distribution of dust in comet 1P/Halley (e.g., Mazets et al., 1986).

A fundamental difference between the polarization phase curves of the gaseous-emission explanation and the weakly absorbing dust, which explains the low $P_{max}$ in comets, appears in the degree of linear polarization near backscattering at α < 30° While the gaseous-emission explanation would result in the absence of the negative-polarization branch, the two-component framework suggests a very deep negative polarization branch, whose amplitude is $|P_{min}| \approx 5–7\%$. Unfortunately, none of the previous comets having such a low $P_{max}$ was observed at small phase angles. Therefore, there is a great deal of interest in future polarimetric measurements of linear polarization in the low-$P_{max}$ comets at α = 10°–15°, where the minimum of the negative polarization is often observed.

## Acknowledgment

E.Z., E.C., and A.K. acknowledge support from the FEFU priority project "Materials". I.L. thanks the SAIA Programme for financial support. The authors thank two anonymous referees for their constructive comments on this manuscript.